# Vertex Detector Cable Considerations

W. E. Cooper

Fermi National Accelerator Laboratory
P. O. Box 500, Batavia, Illinois 60510, USA

Vertex detector cable requirements are considered within the context of the SiD concept. Cable material should be limited so that the number of radiation lengths represented is consistent with the material budget. In order to take advantage of the proposed accelerator beam structure and allow cooling by flow of dry gas, "pulsed power" is assumed. Potential approaches to power distribution, cable paths, and cable design for operation in a 5 T magnetic field are described.

## 1  Introduction

Vertex detector cable details are strongly dependent upon power requirements: hence sensor technology and sensor readout electronics. No definite choices have been made for either of those. To allow progress on a few potential issues and offer guidance, power delivery aspects of cabling are considered within the context of the SiD detector concept. Issues and conclusions will need to be re-evaluated once sensor and readout choices have been made.

To minimize the production of secondary particles and multiple scattering contributions to detector resolution, most vertex detector designs have sought to limit contributions of support structures and sensors to < 0.1% of a radiation length per layer at normal incidence. Cable contributions should not add significantly to the 0.1% of a radiation length budget.

The vertex detector for the SiD concept comprises a short central pixel barrel of length 125 mm and outer radius 60 mm and four inner pixel disks at each end of the barrel, as shown in Figure 1. The inner pixel disks extend coverage for forward tracks as coverage of each barrel layer is lost. Three additional pixel disks per end, possibly with a coarser pixel granularity, supplement the outer tracker and extend forward tracking to $\cos(\theta) = 0.99$.

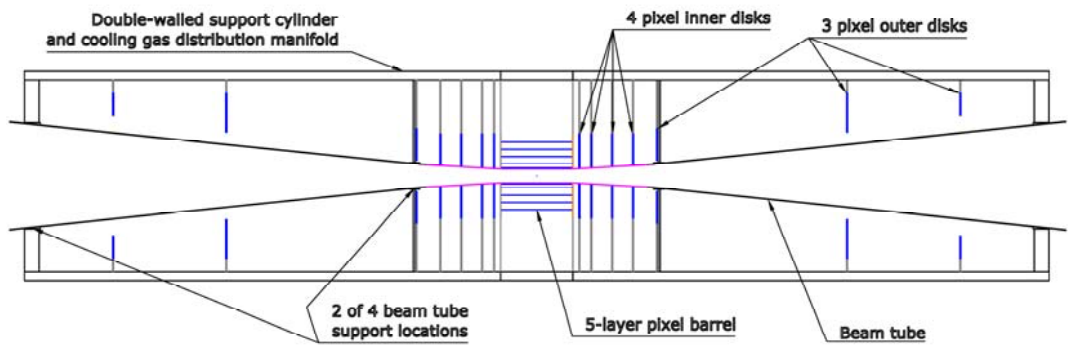

**Figure 1:** Vertex detector elements and support structure

While disk cables can remain outside the vertex detector fiducial volume, it should be clear that portions of barrel cables are within the fiducial volume and warrant special attention.

## 2 Barrel cable paths

Cables from the barrel ends can initially run radially inward, radially outward, or follow paths dictated by layer radius. If cables run radially inward, they can exit the detector fiducial volume quickly and follow the beam pipe surface. However, additional clearance for cables would be needed between disk inner edges and the beam pipe, limiting disk forward coverage. To avoid that issue, this initial analysis assumes that cables run radially outward as shown in Figure 2. A possible location for power distribution cards, should they be needed, is shown as well. This arrangement has the advantage that cables can be dressed and the vertex detector can be tested before installation about the final beam pipe; once testing has been completed, cables and connections need not be disturbed.

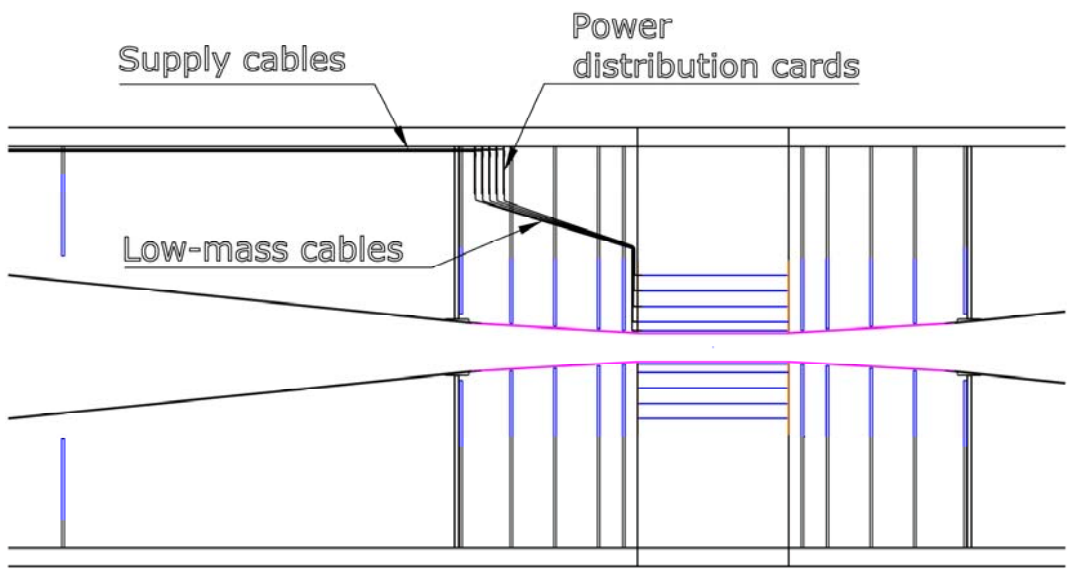

**Figure 2:** One possible path for barrel cables and a location for power distribution cards

## 3 Power, conductor sizing, and magnetic forces

The SiD power dissipation budget is based upon dry gas (nitrogen or air) cooling and a reduction of average power with respect to peak power by a factor of 1/80. For an average power density of 131 µW per $mm^2$ of sensor active area, the present SiD barrel dissipates 21.185 W average and 1695 W peak power. (These values have been updated from those in the LCWS 2008 talk). Small differences between sensors in layer 1 and in layers 2-5 are ignored. If power is delivered via one cable at each end of a sensor R-ϕ location and there are 108 R-ϕ locations, then the average sensor power provided by a cable is 98.1 mW. For this analysis, we assume that 2.5 V must be available at the sensors and 2.9 V is available at the output of power distributors, that is, 0.2 V drop occurs in each supply and each return path of the low-mass cables. During peak power conditions, current per cable is 3.14 A. From that, plus a low-mass cable length of 300 mm (conservatively long), we can deduce the required conductor cross-section. For aluminum conductor with a resistivity of 2.8 μΩ-cm and an available width for the conductor of 5.6 mm (based upon vertex detector sensor width, but conservatively narrow), the required conductor thickness would be 23.5 μm. That is consistent with relatively standard aluminum on kapton flat-line and corresponds to an aluminum thickness slightly less than "1/4 ounce".

The radial run of low-mass cables in front of the innermost disk has a length of about 70 mm and is perpendicular to the magnetic field. For a two-conductor-layer flat-line with 75 μm insulation between conductor layers, the torque due to a 5 T magnetic field is 0.079 N-mm or 7.9 g-mm. For an 8 mm wide cable, that corresponds to forces at cable edges of ±0.5 g. For many applications, that would not be a problem. However, for an application in which power is cycled at a 5 Hz rate, structures are designed to support sensors weighing 0.27 g each, and structural material has been minimized, vibration becomes an issue. That issue is almost entirely eliminated if power is delivered via cables with 3 conductor layers, as suggested in Figure 3. Except for Hall effect corrections and imperfections in current sharing, cable runs would then see zero net force and torque if the center layer were used to supply current and the two outer layers were used to return current.

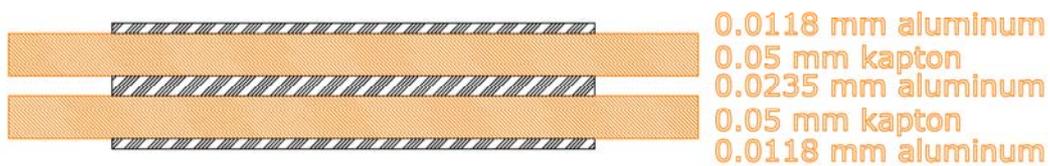

**Figure 3:** Three-conductor-layer power delivery cable; the vertical scale is 10x the horizontal scale.

Within its aluminized region, this cable represents 0.09% of a radiation length at normal incidence. We note that DC-DC conversion or series connections at the power distributors would not diminish current requirements in the final, low-mass cables.

One fault condition deserves special attention. If a common ground at the barrel were provided for all sensors and one sensor were not powered, then supply and return currents in cables of that sensor would not balance. Instead, each of its two cables would see an appropriate share of the total barrel return current: approximately (678 A)(107/108)/(216 cables) = 3.11 A. The force exerted on a radial cable run of length 70 mm would = 1.09 N (equivalent to 109 grams). Though that force is applied at a 5 Hz rate and half of the force could be transferred to cable restraints, the remaining half would be applied to the barrel end and is likely to be sufficient to damage a barrel of low-mass construction. This appears to be a strong argument for considering grounding, power isolation, and failure modes carefully.